\documentclass[%
showpacs,preprintnumbers,
 amsmath,amssymb,
 aps,
 prl,
 lengthcheck,
letterpaper
]{revtex4-1}

\usepackage{graphicx}
\usepackage{hyperref}
\usepackage{ulem}
\usepackage{txfonts}

\newcommand{\ket}[1]{\ensuremath{|{#1}\rangle}}

\begin{document}

\title{Stimulated Neutrino Transformation Through Turbulence}

\author{Kelly M.\ Patton}
\email{kmpatton@ncsu.edu}
\affiliation{Department of Physics, North Carolina State University,
Raleigh, North Carolina 27695-8202, USA}

\author{James P.\ Kneller}
\email{jim\_kneller@ncsu.edu}
\affiliation{Department of Physics, North Carolina State University,
Raleigh, North Carolina 27695-8202, USA}

\author{Gail C.\ McLaughlin}
\email{gail\_mclaughlin@ncsu.edu}
\affiliation{Department of Physics, North Carolina State University,
Raleigh, North Carolina 27695-8202, USA}

\begin{abstract}
We derive an analytical solution for the flavor evolution of a neutrino through
a turbulent density profile which is found to accurately predict the amplitude and transition wavelength of
numerical solutions on a case-by-case basis. The evolution is seen to strongly depend upon those Fourier modes in the
turbulence which are approximately the same as the splitting between neutrino eigenvalues.  Transitions are strongly enhanced by those Fourier modes in the turbulence which are approximately the same as the splitting between neutrino eigenvalues.  We also find a suppression of transitions due to the long wavelength modes when the ratio of their amplitude and the wavenumber is of order, or greater than, the first root of the Bessel function $J_0$.
\end{abstract}

\pacs{14.60.Pq}
\date{\today}

\maketitle

%
%
%

\noindent
\section{Introduction}

The propagation of neutrinos and their associated flavor transformation is a
fascinating problem with many applications. Often neutrinos are propagating 
through environments
where the background potential is not smooth. Examples include propagation 
through the earth, the sun, supernovae,
hypernovae, black hole accretion disks, compact object mergers, and the early universe.  
Fluctuations can occur on many scales in these complex environments, and it is important to understand which density fluctuation scales
will influence neutrino flavor transformation.  

There is a long history of research into MSW-type oscillations due to the small scale fluctuations in the earth and sun  \cite{2001EPJC...20..507O,PhysRevD.43.2484, Nunokawa, Burgess, Sawyer}, as well as the large, even turbulent fluctuations in supernovae \cite{Loreti:1995ae, Friedland:2006ta, Fogli, Kneller:2010sc, Boriello, Choubey,  Kneller:2012id}.  In a supernova environment, neutrino-neutrino interactions must also be considered \cite{Qian:1995, Fuller:2006, Duan:2006, Balantekin:2007, Duan:2007, Duan:2006PRL, Duan:2010, Gava:2009, Esteban-Pretel, Banarjee:2011}. Recently, the effects of density fluctuations on collective neutrino oscillations have been studied \cite{Reid, Cherry, Cherry2}.  However, in order to fully understand these effects, turbulence and the fluctuation scales which are most important the various types of neutrino oscillations must be identified.

In many of the previous works on neutrino propagation through density fluctuations, either the probability distribution of an ensemble of neutrinos  or moments of the distribution \cite{Friedland:2006ta,Kneller:2010sc,2013arXiv1302.3825K,Loreti:1995ae,Fogli}  have been studied.  An ensemble consists of neutrinos traveling through many different realizations of turbulence. Under certain conditions such an ensemble will become fully depolarized \cite{Loreti:1994ry}.  In general for N flavors the depolarization limit is where the final distribution of transition probabilities $P$ is
proportional to $(1-P)^{N-2}$ with a mean of 1/N \cite{Kneller:2010ky}. 

However, it is not always appropriate to use a completely depolarized distribution to characterize the survival probabilities of neutrinos.   Depending on the
typical history of the neutrinos, it may be better to describe the  ensemble with a different distribution, to consider a subset of a distribution through the use of correlations,  or to consider individual neutrinos. 

Whether considering individual neutrinos, constructing the probability distribution of an ensemble, or computing correlations,  the relevant physics is the type and strength of transitions that a neutrino will undergo.

The description of density fluctuations is often a Fourier decomposition into
sinusoids, which will produce arbitrarily complicated density distributions
depending on the number of modes used. The simplest case is that of a single
sinusoidal perturbation super-imposed upon a constant density fluctuation. 
Neutrinos traveling through this potential can exhibit transitions between
states called  parametric resonances \cite{Ermilova,1987PhLB..185..417S,Akhmedov,1989PhLB..226..341K,2009PhLB..675...69K}. Parametric resonances are distinct from 
MSW-type transformations and occur not only when the scale of the sinusoidal fluctuation corresponds to the neutrino mass splitting, but also when it
corresponds to harmonics of the mass splitting \cite{Kneller:2012id}. 
A true understanding of which conditions will cause parametric type transitions requires an analytic description of the survival probability of individual neutrinos going through turbulent profiles.  In this paper we solve this problem.

We consider neutrino flavor propagation through complicated density
profiles, testing examples of up to fifty sinusoids. We first present a numerical flavor transformation calculation through such a profile, then derive an
analytic solution for the flavor transformation probability.  Comparing this expression to the numerical flavor transformation calculations, we demonstrate that the analytic expression effectively predicts,
on a case by case basis, neutrino flavor transformation probabilities as a
function of distance. This expression also makes clear the most important
density perturbation scales. Not only are the perturbations with wavelengths
that correspond to the neutrino mass splitting important, there are additional
longer wavelength modes that, if present, can suppress the parametric resonance
transitions. This second scale is only weakly dependent on the other scales in the problem, such as $\delta m^{2}$ and the energy of the neutrino.  In fact, it depends most strongly on the amplitude of the perturbation. Thus, we show that for any problem involving neutrino flavor transformation
through a medium, it is necessary to understand density fluctuations on two
length scales: the scale of the neutrino mass splitting 
$ \lambda_{fluct, split} \sim 20\, {\rm km} \, \left[ \left( \left(\frac{\delta m^{2}}{3\times10^{-3}\,{\rm eV^{2}}}\right)\,\left(\frac{20 {\rm MeV}}{E}\right)\, \left(\frac{\cos 2\theta}{0.95}\right) - 0.53 \left( \frac{\rho}{1000 \, {\rm g/cm^{3}}}\right) \right)^{2} \right.\\
\left. + \, 0.1\left(\left(\frac{\delta m^{2}}{3\times10^{-3}\,{\rm eV^{2}}}\right)\,\left(\frac{20 {\rm MeV}}{E}\right)\, \left(\frac{\sin 2\theta}{0.3}\right)\right)^{2} \right]^{-1/2}$, 
and the scale that corresponds to the amplitude of the density fluctuations
$\lambda_{fluct, ampl} \sim 800\, {\rm km} \, \left( \frac{0.1}{C}\right) \, \left( \frac{1000 \, {\rm g/cm^{3}}}{\rho}\right)$. 
\\

\noindent
\section{A Numerical Solution}

The physical quantity we aim to calculate is the probability that an initial
neutrino state $\ket{\nu(r)}$ at $r$ is later detected as the
state $\ket{\nu'(r')}$ at $r'$. This probability can be computed from the
$S$-matrix which relates the initial and final neutrino states by the 
equation $\ket{\nu'(r')} = S(r',r)\,\ket{\nu(r)}$ \cite{Kneller:2006, Kneller:2009}. The $S$-matrix evolves
according to the differential equation 
\begin{equation}
\imath \frac{dS}{dr} = H\,S \label{eq:dSdx}
\end{equation}
where $H$ is the Hamiltonian. In the flavor basis $H$ is given by
$H^{(f)} = U_{0}\,K^{(m)}_{0}\,U_{0}^{\dagger} + V^{(f)}$
where $U_0$ is the vacuum mixing matrix, $K^{(m)}_{0}$ is the diagonal matrix
of vacuum eigenvalues and $V^{(f)}$ some `potential'. We shall assume the
potential $V$ possesses a `smooth' component, which we denote by $\breve{V}$,
and a `perturbation' $\delta V$. With this assumption we can regroup the Hamiltonian into $H = \breve{H} + \delta
V$. We now enforce the requirement that the potential $V$ is `MSW' like in the sense that the only non-zero entry of $V^{(f)}$ is $V_{ee}$ and that the component $\breve{V}_{ee}$ is a
constant, $V_0$.  We now focus upon the perturbation and consider the case where $\delta V$ is
built from a series of $N_k$ sinusoidal fluctuations with wavenumbers $q_j$, amplitudes
$C_j$ and phase shifts $\eta_j$. Like $\breve{V}$, we restrict ourselves to the case where only $\delta V_{ee}$ is non-zero. 
In full the perturbation is written as 
\begin{equation}
\delta V_{ee}(r) = V_0 \left\{ \sum_{a=1}^{N_k} C_a \sin\left(q_a r+\eta_a\right) \right\}
\end{equation}   
Turbulence is often represented as a Fourier series of exactly this form with
realizations generated by assigning random values for the amplitudes,
wavenumbers and phase factors according to some algorithm. 
The phases $\eta_a$ are typically uniformly distributed from zero to $2\pi$ but the random amplitudes and 
wavenumbers are generated using algorithms that are functions of the power spectrum.
Two commonly considered cases are the white-noise power spectrum and the
case where the power spectrum is an inverse power law.
\begin{figure}
\includegraphics[clip,width=\linewidth]{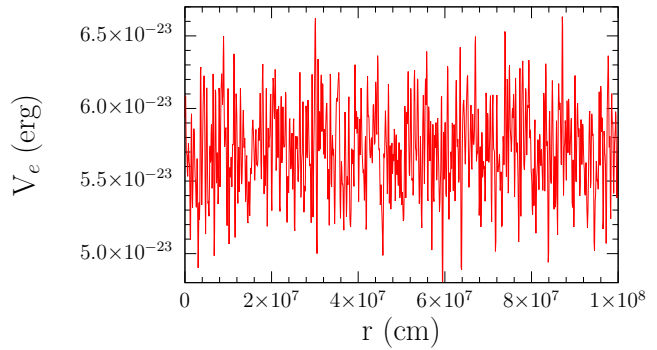}
\caption{A turbulent MSW potential. The mean $V_0$ of the potential is set
to be one half the MSW resonance for a $20\;{\rm MeV}$ neutrino with 
mixing angle $\sin^2 2\theta = 0.1$ and mass splitting $\delta m^{2} = 3\times
10^{-3} \;{\rm eV^2}$. The turbulence is composed of $N_k = 50$ sinusoids using 
the algorithm described in the text.} \label{fig:1}
\end{figure}
\begin{figure}
\includegraphics[clip,width=\linewidth]{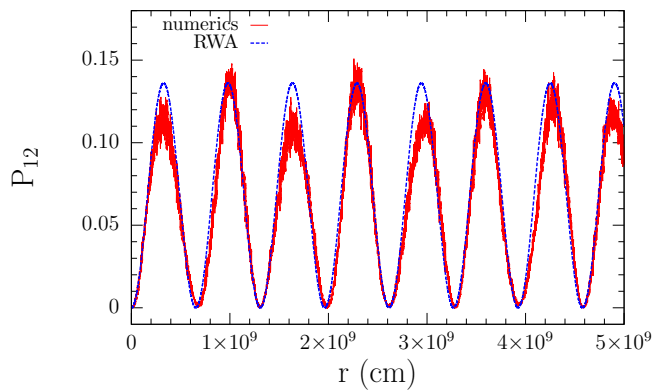}
\caption{The numerical solution for the transition probability between the
unperturbed neutrino states through the potential shown in figure
(\ref{fig:1}). 
The sinusoidal curve marked `RWA' is an analytical prediction for the
evolution.} \label{fig:2}
\end{figure}
In figure (\ref{fig:1}) we show a realization of the white-noise case using $N_k=50$ where the wavenumbers are uniformly distributed between $0.0045\;{\rm km}^{-1}$ and $2.2\;{\rm km}^{-1}$ and 
the amplitudes are uniformly distributed between 0 and 0.02.
The probability we report is the transition probability between the eigenstates of $\breve{H}$. The eigenbasis of
$\breve{H}$ - which we call the unperturbed matter basis, denoted by superscript $(\breve{m})$ - is related to the flavor basis by the 
unitary mixing matrix $\breve{U}$ defined by the requirement that it diagonalize
$\breve{H}^{(f)}$ i.e.\ $\breve{H}^{(f)} = \breve{U}\,\breve{K}^{(\breve{m})}\,\breve{U}^{\dagger}$ 
where $\breve{K}^{(\breve{m})}$ is the diagonal matrix of eigenvalues,
$\breve{k}_{1}, \breve{k}_{2},\ldots$, of $\breve{H}$. Note that the unperturbed matter basis is not equivalent to the instantaneous matter basis.  The unperturbed matter basis includes only the 'smooth' potential $\breve{V}$, and not the perturbation $\delta V$. The unperturbed matter basis is the most suitable basis to calculate the transition probability because transitions vanish when the perturbation vanishes.

The numerical solution for the transition probability for a two-flavor neutrino (we use two-flavors only for the sake of simplicity) with an energy of $20\;{\rm
MeV}$, vacuum mixing angle of $\sin^2 2\theta = 0.1$ and mass splitting $\delta m^{2} = 3\times 10^{-3} \;{\rm eV^2}$, through the turbulence in figure (\ref{fig:1}) can be observed 
in figure (\ref{fig:2}). The mean $V_0$ of the potential in figure (\ref{fig:1}) was set to be one half the MSW resonance potential for this energy and mixing parameters. 
The evolution of $P_{12}$ is really quite remarkable: given the density profile shown in figure (\ref{fig:1}) one might expect that either the solution for $P_{12}$ would drift gradually 
to $P_{12} =0.5$ or that $P_{12}$ would evolve as chaotically as the potential. The actual evolution of $P_{12}$ does neither of these things: it is quasi periodic containing 
a prominent sinusoidal mode. The curve marked `RWA' in the figure is a prediction for this component that we now describe.\\

\noindent
\section{Predictions for the Wavelength and Amplitude}

Our analytical treatment for the effect of multiple sinusoidal density
fluctuations is a generalization of the treatment presented in Kneller, McLaughlin \& Patton
\cite{Kneller:2012id} for a single sinusoid. 
Although it is possible to solve this problem with arbitrary numbers of neutrino
flavors, we shall restrict ourselves to just two, again, for the sake of clarity.  Our first step is to transform the problem into the unperturbed matter basis using the mixing matrix $\breve{U}$.  After this transformation, the Hamiltonian is written as 
\begin{equation}
H^{(\breve{m})} = \breve{K}^{(\breve{m})} - \imath
\breve{U}^{\dagger}\,\frac{d\breve{U}}{dr} +
\breve{U}^{\dagger}\delta V^{(f)}\breve{U}.
\end{equation} 

We write the $S$-matrix for the unperturbed matter basis as the product
$S^{(\breve{m})} = \breve{S}\,A$ where $\breve{S}$ is solution of the
unperturbed problem i.e. the constant density Hamiltonian $\breve{H}$, defined by the equation
\begin{equation}
\imath \frac{d\breve{S}}{dr} = \left[ \breve{K}^{(\breve{m})} - \imath
\breve{U}^{\dagger}\,\frac{d\breve{U}}{dr} \right] \,\breve{S}.
\end{equation}   
Note that we have suppressed the factors of $\hbar c$ that occur in the
S-matrix equation.

For this potential $\breve{S}$ is simply $\breve{S} =\exp\left(-\imath\breve{K}^{(\breve{m})}\,x\right)$ because the eigenvalues of
$\breve{H}$ are constant.
Since $\breve{S}$ is known we can solve for the effect of the perturbation $\delta V$ by finding the solution for $A$, given by the differential equation 
\begin{equation}
\imath \frac{dA}{dr} = \breve{S}^{\dagger}\,\breve{U}^{\dagger}\delta V^{(f)}\breve{U}\,\breve{S} \,A. \label{dAdr}
\end{equation}

The Hamiltonian for $A$ possesses both diagonal and off-diagonal elements but those diagonal elements can
be removed by writing the matrix $A$ as $A=W\,B$ where $W=\exp(-\imath\Xi)$ and
$\Xi$ is a diagonal matrix $\Xi=diag(\xi_{1},\xi_{2})$, see \cite{Kneller:2012id}.  Using equation (\ref{dAdr}), we find the differential equation for $B$ is 
\begin{equation}
\imath \frac{dB}{dr} = W^{\dagger}\left[\breve{S}^{\dagger}\breve{U}^{\dagger}\delta V^{(f)}\breve{U}\,\breve{S} -\frac{d\Xi}{dr}\right]\,W\,B,
\end{equation}
where the diagonal matrix $\Xi$ is chosen in order to remove the diagonal elements of $\breve{S}^{\dagger}\breve{U}^{\dagger}\delta V^{(f)}\breve{U}\,\breve{S}$. 

The quantities $\xi_{i}$ using our form for $\delta V$ are 
\begin{equation}
\xi_{i} = V_0\,|\breve{U}_{ei}|^{2} \left\{ \sum_{a=1}^{N_k}
\frac{C_a}{q_a}\left[\cos\eta_a - \cos\left(q_a
r+\eta_a\right)\right] \right\}. 
\end{equation}
Using this result, the equation for $dB/dx$ is found to be
\begin{equation}\label{eq:idBdr}
\begin{split}
\imath \frac{dB}{dx} = V_0 \left\{ \sum_{a=1}^{N_k} C_
a \,\sin\left(q_a r+\eta_a\right) \right\} \\
\left( \begin{array}{cc}  0 & \breve{U}_{e1}^{\star}\breve{U}_{e2}
e^{\imath\left(\delta\breve{k}_{12} r+\delta\xi_{12}\right)} \\
      \breve{U}_{e2}^{\star}\breve{U}_{e1}e^{-\imath\left(\delta\breve{k}_{12}
r+\delta\xi_{12}\right)} & 0 
     \end{array} \right)\,B   
\end{split}
\end{equation}
where $\delta\breve{k}_{12} = \breve{k}_{1} -\breve{k}_{2}$ and similarly
$\delta\xi_{12}=\xi_{1}-\xi_{2}$. 
The next step is to use the Jacobi-Anger expansion for the complex exponentials
$\exp\left(\imath\delta \xi_{12}\right)$
\begin{eqnarray}\label{eq:JAexpansion}
\exp\left(\imath\delta \xi_{12}\right)&& = \prod_{a=1}^{N_k} 
\sum_{n_a=-\infty}^{\infty} (-\imath)^{n_a} J_{n_a}\left(z_{a}\right) \nonumber
\\
&& \times \exp\left[\imath z_{a}\cos\eta_a + \imath\,n_a\left(q_{
a}\,r+\eta_a\right)\right] 
\end{eqnarray}
where $J_{n}$ is the Bessel J function and, in order to tidy up the notation, we
have introduced the quantities $z_{a}$ defined to be
\begin{equation}\label{eq:zDef}
z_{a} = \frac{C_a V_0}{\hbar c q_a}
\left(|\breve{U}_{e1}|^{2}-|\breve{U}_{e2}|^{2}\right). \\
\end{equation}
In Eq. \ref{eq:zDef} we have added back in the factor of $\hbar c$. The Jacobi-Anger expansion is really a Fourier series expansion, where each term has a coefficient $(-\imath)^{n_a} J_{n_a}\left(z_{a}\right)  \exp\left[\imath z_{a}\cos\eta_a\right].$  For $n=0$, this coefficient is simply the mean value of the complex exponential.

Using this expansion and the definition we find
\begin{equation}\label{eq:dBdr:offres} 
\imath \frac{dB}{dx} = 
\left( \begin{array}{cc}
     0 & h_{12} \\
     h_{21} & 0
    \end{array}\right)\,B  
\end{equation} 
where the elements $h_{12}$ and $h_{21}$ are given by 
\begin{equation}\label{eq:hij}
\begin{split}
h_{12} = h_{21}^{\star} =
\frac{\breve{U}_{e1}^{\star}\breve{U}_{e2}}{|\breve{U}_{e1}|^2 -|\breve{U}_{e2}|^2} 
\sum_{a=1}^{N_k}\,\sum_{n_a=-\infty}^{+\infty}\, n_a\,q_{
a}\,\kappa_{a,n_a} \\
\prod_{b=1,b\neq a}^{N_k}\left\{\sum_{n_b=-\infty}^{+\infty}
\kappa_{b,n_b}\,\exp\left[\imath(\delta\breve{k}_{12} + n_a\,q_{ a}+
n_b\,q_{ b}) r\right] \right\}
\end{split}
\end{equation} 
with the complex parameters $\kappa_{a,n_a}$ defined to be 
\begin{equation}
\kappa_{a,n_a} = (-\imath)^{n_a}\,J_{n_a}(z_{a})
\,\exp\left[\imath\left(n_a\,\eta_a +z_{a} \cos\eta_a \right)\right].
\end{equation} 
Note that part of the definition of $\kappa$ is the coefficient from the Jacobi-Anger expansion.  This form for the components of the Hamiltonian closely matches that found for similar systems in molecular physics, such as that studied in Kondo, Blokker, \& Meath \cite{Kondo:1992}.

To make additional progress we make use of the Rotating Wave
Approximation (RWA) and drop all terms in each infinite series - the $n_a$'s and
$n_b$'s - in equation (\ref{eq:hij}) except the most `important'.
The most important set of integers can be found from the criterion used for the
case of a single sinusoidal perturbation. 
We select the values for the $n_a$'s which come closest to satisfying the
parametric resonance condition $|\delta\breve{k}_{12} + \sum_{a} n_a q_{ a}| \approx 0$. We denote these
values by $n_{\star a}$. The RWA removes the sum over each $n_a$ and $n_b$ and we can define the quantity 
\begin{equation}\label{eq:kappaDef}
\kappa = \frac{\breve{U}_{e1}^{\star} \breve{U}_{e2}}{|\breve{U}_{e1}|^2 - |\breve{U}_{e2}|^2}
\left(\sum_{a=1}^{N_k} n_{\star a}\,q_{ a} \right)
\prod_{a=1,}^{N_k}\kappa_{a,n_{\star a}} \\
\end{equation}
After making the RWA we can write out the solution for $B$ after introducing $2p = \delta\breve{k}_{12} + \sum_{a} n_{\star a} q_{ a}$ and $Q^{2} = p^{2}+ \kappa^{2}$, see Kneller, McLaughlin \& Patton \cite{Kneller:2012id}. 

The RWA allows us to write the differential equation for $B$ as 
\begin{equation}
\imath \frac{dB}{dr} = H^{(B)}\,B, 
\end{equation} 
and define the RWA Hamiltonian as 
\begin{equation}
H^{(B)} = 
\left( \begin{array}{cc}
     0 & \kappa \exp{(2\, \imath\, p\, r)}\\
      \kappa^{\star} \exp{(-2\, \imath \, p\, r)} & 0
    \end{array}\right)\, .
\end{equation} 
Following the procedure outlined in \cite{Kneller:2012id}, we find the solution for $B$ is
\begin{equation}
\begin{split}
& B = \\ 
& \left( \begin{array}{cc}
 e^{\imath p r} \left[ \cos(Q r) -\imath\frac{p}{Q}\sin(Q r)
\right] & -\imath e^{\imath p r}\,\frac{\kappa}{Q}\sin(Q r) \\
-\imath e^{-\imath p r}\frac{\kappa^{\star}}{Q}\sin(Q r) & e^{-\imath
p r} \left[ \cos(Q r) +\imath\frac{p}{Q}\sin(Q r) \right]
 \end{array} \right).
 \end{split}
\end{equation}

Since we assumed the unperturbed potential was a constant, the unperturbed matrix
$\breve{S}$ and the matrix $W$ are diagonal matrices, so we find 
that the transition probability between the unperturbed matter states 1 and 2 is of the form  
\begin{equation}
P_{12} = |B_{12}|^{2}= \frac{\kappa^2}{Q^2}\,\sin^{2}(Q r). \label{eq:P_12}
\end{equation}
In practice we found that when the number of sinusoids $N_k$ is large it is computationally prohibitive
to locate the values of the $N_k$ integers which minimize the phase $|\delta k_{12} + \sum_a n_a q_{ a}|$ via a scan. 
When $N_k$ is large our strategy for locating the RWA solution is to use $-\kappa^{2}/Q^{2}$, or the negative of the amplitude, 
as a `potential' in the $N_k$ dimensional space of integers and then locate the minimum of the potential using a simplex. 
The RWA solution for the turbulence shown in figure (\ref{fig:1}) - using the simplex algorithm to locate 
the integers - is the second solution shown in figure (\ref{fig:2}) marked `RWA'. The reader will observe
that it is a good description of both the amplitude and the wavelength of the dominant sinusoidal component of $P_{12}$. \\

\noindent
\section{Suppressed Transitions}

\begin{figure}
\includegraphics[clip,width=\linewidth]{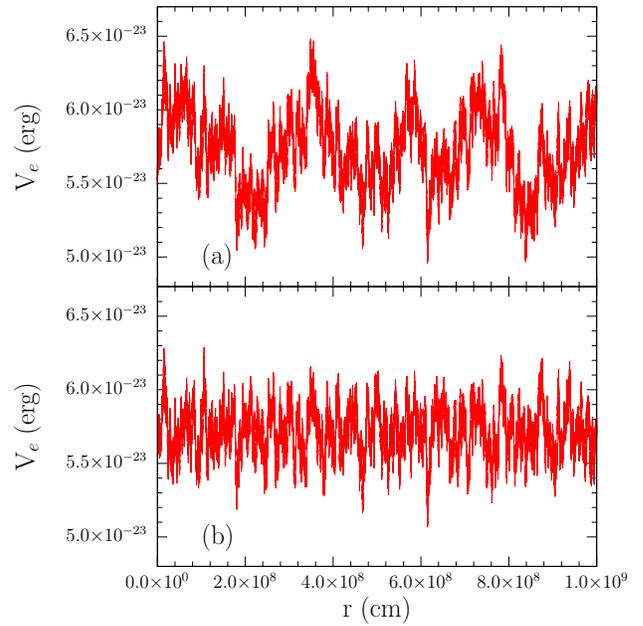}
\caption{Two turbulent neutrino potentials. In the upper panel the realization
is uses an inverse power law spectrum. 
The lower panel is the exact same set of amplitudes, wavenumbers and phases as
the upper panel but with the five lowest wavenumbers removed.} \label{fig:3}
\end{figure}
\begin{figure}
\includegraphics[clip,width=\linewidth]{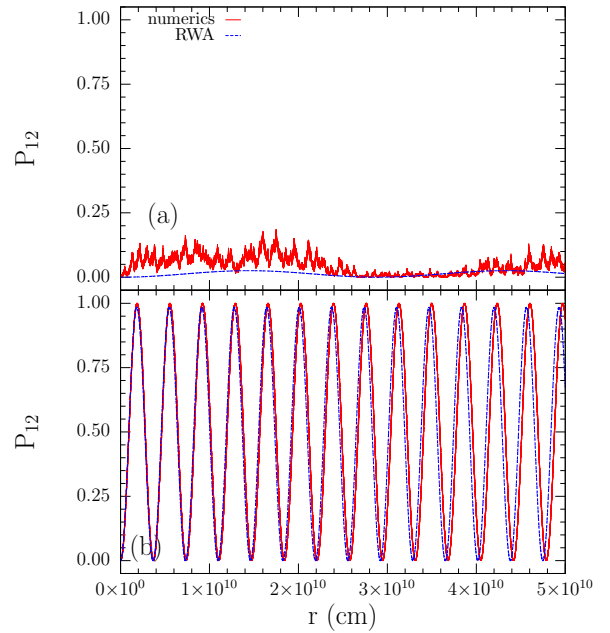}
\caption{The numerical and analytical solutions for the neutrino evolution
through the turbulent potentials shown in figure (\ref{fig:3}).} \label{fig:4}
\end{figure}

The success of the RWA solution shown in figure (\ref{fig:2})
indicates the theory describes the overall features of the numerical solution well. The RWA solution obviously depends upon
those wavenumbers with non-zero contributions to fulfilling the parametric resonance condition
$\delta\breve{k}_{12} + \sum_{a} n_a q_{ a} \approx 0$.  Defining $\lambda_{fluct,split} \sim 2\pi/q_{ a}$ for these modes, we find the scale given in the introduction. But that is \emph{not} the whole story: the quantity $\kappa$, which controls the amplitude and the wavelength of the transitions, actually depends upon \emph{every}
wavenumber even if the RWA solution indicates it does so with $n_{\star a} =0$. 

To understand this effect, consider equation (\ref{eq:idBdr}).  We can break the matrix elements of the Hamiltonian in this equation into three parts.  The coefficient of the exponential, $\breve{U}_{e1}^{\star} \breve{U}_{e2} V_0 \left\{ \sum_{a=1}^{N_k} C_a\,\sin\left(q_{ a}r+\eta_a\right) \right\}$, or the complex conjugate, is an operator which connects the two neutrino states.  The complex exponential $e^{\imath \breve{\delta k}_{12} r}$ is the unperturbed solution to the problem and it describes the evolution of the neutrinos in the constant density potential.   Finally, the complex exponential $e^{\imath \delta \xi_{12}}$ is the distortion to the unperturbed solution by the diagonal elements of $\delta V$.  This distortion reflects the modification of oscillation frequency of the unperturbed system. As we discuss in the following paragraphs, the modes which cause the parametric resonance transitions create little change in this frequency, but longer wavelength modes can create larger changes.

This term can be written as a product of the distortions from each individual mode:
\begin{equation}
e^{\imath \delta\xi_{12}} = \prod_{a=1}^{N_{k}} e^{\imath \delta\xi_{12 a}},
\label{eq:distortion}
\end{equation}

Recall that in deriving the solution \ref{eq:P_12} we expanded each $e^{\imath \delta\xi_{12 a}}$ using the Jacobi-Anger expansion \ref{eq:JAexpansion}, and retained only the relevant $n_{\star a}$ term. For the modes which don't contribute to the parametric resonance, this is the $n_{\star a} = 0$ term.  Through an identity of Bessel functions this term can be written as 
\begin{equation}\label{eq:distortion}
 J_{0}\left(z_{a}\right)  \exp\left[\imath z_{a}\cos\eta_a\right] = \langle e^{\imath \delta \xi_{12 a}} \rangle,
\end{equation}
 where the average value is defined explicitly as
\begin{equation}\label{eq:aveDistortion}
\langle e^{\imath \delta \xi_{12 a}} \rangle = \frac{2\pi}{q_{ a}} \int_{0}^{2\pi/q_{ a}}{dr \,e^{\imath \delta \xi_{12 a}}}.
\end{equation}
Thus for all the modes that don't contribute to the parametric resonance, each distortion can be represented by an average value and the oscillatory terms in the Jacobi-Anger expansion in equation (\ref{eq:JAexpansion}) can be neglected. As long as the average value of the distortion, $J_0(z_{a})$, for all these modes is close to unity, then these modes which don't contribute to the parametric resonance make only small modifications to the amplitude and wavelength of the solution \ref{eq:P_12}. But when the average value of the distortion is large, i.e.\ $|J_0(z_{a})| < 1$, then these modes can significantly modify they amplitude and wavelength of the solution even though the do not contribute to the parametric resonance!

Examining the argument of the Bessel function, $z_{a} \sim C_a V_0/(\hbar c q_{a})$, we see that the criteria for determining whether a mode causes significant distortion is the 
ratio of its amplitude to its wavenumber. As long as the fluctuation amplitude
$C_{a} V_0/ (\hbar c) $ is much smaller than the wavenumber $q_{a}$, then $\langle e^{\imath \delta \xi_{12 a}} \rangle \sim 1$, and that mode will cause little distortion. However, if $z_{a} \sim C_{a} V_0/(\hbar c q_{a}) > 1$ then the value of the Bessel function will be small and there will be considerable distortion, $\langle e^{\imath \delta \xi_{12 a}} \rangle < 1$. In addition, if there is a mode for which the value of $z_{a}$ hits a zero of a Bessel function then there is maximal distortion $\langle e^{\imath \delta \xi_{12 a}} \rangle \sim 0$. In these situations, the transitions are extraordinarily strongly suppressed since the distortion of all the modes appears as a multiplicative factor.

We can identify the largest wavenumbers which cause significant distortion by examining 
the point where $J_{0}(z_{a})$ hits its first zero, the first of which occurs when $z_{a} \sim 2.4$.  Using equation (\ref{eq:zDef}), we find the wavenumbers of interest by setting $z_{a} \sim C_{a} V_{0}/(\hbar c q_{a}) \sim 2.4$.  If we define $\lambda_{fluct,ampl} \sim 2\pi/q_{a}$ for these modes, we find the expression given in the introduction. From the scaling of $z_{a}$ with $q_{a}$ we discover it is the long wavelength modes which dominate this effect, not the small wavelengths.

We show this dependence upon the long wavelength turbulence modes by generating the two turbulence density profiles shown in figure (\ref{fig:3}) using an inverse power law spectrum (see \cite{2013arXiv1302.3825K,2013PhRvD..88d5020K,2013PhRvD..88b3008L} 
for a description of the algorithm we use for this case) with the rms amplitude of the turbulence set to 0.05, a spectral index of $\alpha =5/3$, and a long wavelength cutoff set to $\sim 3000\;{\rm km}$. 
Since $V_0$ is only half the MSW density, it is would require an extremely unlikely, $20-\sigma$, fluctuation to approach the MSW density with this turbulence amplitude. 
After generating the turbulence, we slightly increased the wavelength and amplitude of the longest wavelength mode by hand to give $z_{1} \sim 2.4$.  
The only difference between the two profiles is that the five longest wavelength modes from
the top panel have been removed for the lower. The missing modes possess wavenumbers many orders of magnitude 
smaller than the eigenvalue splitting $\delta \breve{k}_{12}$. 
The RWA solution for these realizations are found to be $n_{\star a} =0$ for every mode, including the five longest wavelengths, except for one mode 
which happens to satisfy the parametric resonance condition by itself. 
If there were no dependence upon the wavelengths where $n_{\star a} =0$ then removing the five longest wavelength modes should not change the solution. 
However, when we actually solve for $P_{12}$ with and without these five wavelengths we do find different solutions as shown in figure (\ref{fig:4}). 
Note that the horizontal scale on the figure is more than one hundred times larger than the longest wavelength mode in the turbulence i.e.\ the cutoff.
Without the five longest wavelengths the amplitude of the transition is unity and the wavelength of order $\sim 3 \times 10^{9}\;{\rm cm}$: with the five 
longest wavelengths the amplitude is only $\sim 0.1$ and the wavelength stretches to $\sim 3 \times 10^{10}\;{\rm cm}$. 

Inspection of the values of $z_{a}$ for the 
five modes indicates they all lie in the vicinity of the first root of $J_0$.  The distortion predicted from these five modes together, obtained by multiplying the individual distortions given by equation (\ref{eq:distortion}), is $\sim 0.02$. This is mainly due to the first mode, which had $z_{1} \sim 2.4$, although the other four missing modes also contribute.  Even though these five modes all have $n_{\star a} =0$, they still have a strong effect on the transition.

The example shown in figure (\ref{fig:4}) was specifically tailored to show how dramatic the suppression effect can be.  However, we also found that it occurs in more general examples.  Using a white noise spectrum with wavelengths between 3 km and 300 km and amplitudes up to 0.025, 20 examples with $N_{k} = 20$ where a parametric resonance occurred were found.  We added $N_{k}=20$ extra modes with wavelengths between 300 km and 3000 km, then 300 km to 30000 km, to those original white noise spectra.  This resulted in perturbations with $N_{k} = 40$ that were known to have a parametric resonance, but also had the possibility of exhibiting suppression.  In approximately half of the trials, the amplitudes found in the numeric simulations showed suppression varying from just a few percent to almost 50\%.  From these results, we expect that suppression will occur for approximately half of turbulent spectra, with the degree of suppression depending on the values of wavelength and corresponding amplitudes within the spectrum.  

Finally, we note the analogy, with the Stark effect for a two-level atom.  For a two-level atom, transitions are stimulated between the states by a laser if the laser frequency matches the energy splitting of the system.  Placing the atom in an external electric field causes shifts in the wave functions and energies of the original (unperturbed) states.  Once this shift occurs, a laser that stimulated transitions prior to the application of the electric field can no longer do so.  In the neutrino case, the ``lasers'' are the modes fulfilling the parametric resonance condition, and the external electric field corresponds to modes with $n_{\star a} = 0$.

%
%

\noindent
\section{Conclusions}

Understanding neutrino propagation through turbulent media is essential to understanding future measurements of supernova neutrino spectra.  While previous treatments have focused primarily on the behavior of ensemble averages, we have instead considered the exact propagation of single neutrinos.  Our analytic results show that density perturbations play two roles.  We confirm that density perturbations can cause transition between states.  Even in very complicated density profiles the density fluctuations that occur on wavelengths that match the natural wavelength of the neutrinos cause the biggest transition.  However, we find that longer wavelength density fluctuations also {\it suppress} transitions.  They do so by causing the basis state to significantly fluctuate away from its original state on scales shorter than the parametric resonance transition.

A clear implication of these results is that in any environment where neutrinos propagate, it is important to understand the density fluctuations both on the scale of the effective mass splitting and on the scales where the amplitude of the fluctuation corresponds to its wavenumber.  Together with the results presented in this paper, this information can be used to determine the strength and number of transitions a neutrino will experience as it passes through a turbulent ensemble.  This aids in determining the applicability of the depolarization limit of an ensemble of neutrinos.

Many environments have non-constant base density profiles.  To make useful predictions in environments such as supernovae, extensions beyond the constant base profile must be considered.


\acknowledgments

This work was supported by DOE grants DE-SC0004786 (GCM+JPK+KMP) and DE-SC0006417 (JPK),
DE-FG02-02ER41216 (GCM+KMP) and by 
a NC State University GAANN fellowship (KMP). The authors would also like to thank William Meath for useful discussions.


\end{document}